\documentclass[a4paper,twocolumn,superscriptaddress,11pt]{quantumarticle}
\pdfoutput=1
\usepackage[utf8]{inputenc}
\usepackage[english]{babel}
\usepackage[T1]{fontenc}
\usepackage{amsmath}
\usepackage{hyperref}
\usepackage{color,soul}

\usepackage{tikz}
\usepackage{lipsum}

\newcommand{\ket}[1]{\left\vert#1\right\rangle}
\newcommand{\bra}[1]{\left\langle#1\right\vert}
\newcommand{\comm}[2]{\left[#1,#2\right]}

\newcommand{\syse}{\end{array}\right.}

\newcommand{\ag}[1]{}

\usepackage[numbers,sort&compress]{natbib}

\begin{document}

\title{Spin chains for two-qubit teleportation}
\date{\today}

\author{Tony J. G. Apollaro}
\affiliation{Department of Physics, University of Malta, Msida MSD 2080, Malta}
\email{tony.apollaro@um.edu.mt}

\author{Guilherme M. A. Almeida}
\affiliation{Instituto de F\'{i}sica, Universidade Federal de Alagoas, 57072-900 Macei\'{o}, AL, Brazil}

\author{Salvatore Lorenzo}
\affiliation{Dipartimento di Fisica e Chimica, Universit\'{a}  degli Studi di Palermo, via Archirafi 36, I-90123 Palermo, Italy}

\author{Alessandro Ferraro}
\affiliation{Centre for Theoretical Atomic, Molecular and Optical Physics,
Queen’s University Belfast, Belfast BT7 1NN, United Kingdom}
%\email{....}

\author{Simone Paganelli}
\affiliation{Dipartimento di Scienze Fisiche e Chimiche, Universit\`{a} dell'Aquila, via Vetoio, I-67010 Coppito-L'Aquila, Italy}
%\email{....}

%\homepage{http://quantum-journal.org}
%\orcid{0000-0003-0290-4698}
%\thanks{You can use the \texttt{\textbackslash{}email}, \texttt{\textbackslash{}homepage}, and \texttt{\textbackslash{}thanks} commands to add additional information for the preceding \texttt{\textbackslash{}author}. If applicable, this can also be used to indicate that a work has previously been published in conference proceedings.}

%\author{second author}
%\affiliation{example}
%\email{....}

\maketitle

\begin{abstract}
  Generating high-quality multi-particle entanglement between communicating parties is the primary resource in quantum teleportation protocols. To this aim, we show that the natural dynamics of a single spin chain is able to sustain the generation of two pairs of Bell states ---possibly shared between a sender and a distant receiver--- which can in turn enable two-qubit teleportation. In particular, we address a spin-$\frac{1}{2}$ chain with $XX$ interactions, connecting two pairs of spins located at its boundaries, playing the roles of sender and receiver. In the regime where both end pairs are weakly coupled to the spin chain, it is possible to generate at predefinite times a state that has vanishing infidelity with the product state of two Bell pairs, thereby providing nearly unit fidelity of teleportation. We also derive an effective Hamiltonian via a second-order perturbation approach that faithfully reproduces the dynamics of the full system. 
\end{abstract}

\section{Introduction}\label{S.Intro}

Quantum Information Processing (QIP) has become the subject of an increasingly intensive theoretical and experimental effort over the last few decades. With research fields spanning from computation to simulation and metrology, QIP aims at leading the next quantum revolution by developing QIP devices able to outperform any classical analogue in a variety of tasks, from crypthographic key distribution to simulation of chemical reactions. Nevertheless, a necessary condition for almost any QIP task is the capability of implementing a faithfull Quantum State Transfer (QST) protocol \cite{bose2007,nikolopoulos2013quantum,apollarorev}. Indeed, QIP tasks such as quantum key distribution and quantum computation require the transfer of quantum information from a sender to a receiver, embodied by measurement apparatus or quantum processors.  

The means by which QST is achievable can be grouped in three large classes. The first one involves the physical displacement of a carrier encoding the information (\textit{e.g.}, photons) and have been successfully employed in cavity QED-based architectures \cite{ritter12,kurpiers18}.
The second one relies on the dynamics of a physical quantum channel connecting the sender and the receiver, the former encoding the information in a stationary qubit at its location, with the aim that the evolution of the quantum channel allows the information to be retrieved at the receiver's stationary qubit location. In this context, spin-$\frac{1}{2}$ chains have been intensively investigated as faithful quantum channels
for a variety of tasks \cite{PhysRevLett.91.207901,christandl04, wojcik05,almeida18pra,almeida17pra,kay17quantum} \textcolor{teal}. Finally, the third QST protocol is based on exploiting a preexisting quantum resource, usually entanglement, and perform a teleportation protocol ---which represents the most most prominent example of quantum communication under LOCC (local operations and classical communication) constraints. In this paper, we focus on the use of a spin-$\frac{1}{2}$ chain to generate such a quantum resource, which can then be used for the deterministic teleportation of an arbitrary two-qubit state. 

Since the seminal work by Bennett {\textit{et al.}} that introduced the quantum teleportation protocol of a single qubit via the use of a Bell pair and a classical communication channel~\cite{PhysRevLett.70.1895}, a great effort has been devoted both to its experimental implementation~\cite{PhysRevLett.80.1121} and to the generalization to higher dimensional systems ---in particular, $n$-qubit teleportation protocols. The latter find a natural application in LOCC-constrained quantum communication, where high-dimensional systems guarantee higher security and increased transmission rates \cite{bechmannpasquinucci2000quantum, cerf2002security, bru2002optimal, acin2003security, karimipour2002quantum, durt2004security, nunn2013largealphabet, mower2013highdimensional, lee2014entanglementbased, zhong2015photonefficient}. Also, $n$-qubit teleportation protocols can be used in quantum computation, especially in distributed approaches \cite{beals2013efficient} ---where the state of a quantum register needs to be routed to different processing units--- and in client-server models \cite{qiang2017quantum} ---where quantum computation is performed by a remote unit.

While in the original protocol in Ref.~\cite{PhysRevLett.70.1895} the quantum channel for deterministic teleportation is embodied by one of the Bell states, many other states have been found to achieve the same goal, amongst which three-particle GHZ and a class of W states~\cite{PhysRevA.58.4394,PhysRevA.74.062320}. 
The search for 2-qubit teleportation protocols went along the same line: from the original proposal exploiting tensor products of two Bell states~\cite{PhysRevA.71.032303, PhysRevA.72.036301} to genuine four- and five-qubit entangled states~\cite{PhysRevLett.96.060502, PhysRevA.77.032321} and a class of four-qubit states having cluster states as a special case~\cite{1751-8121-45-40-405303}. Similarly, for $n$-qubit teleportation, $2n$-qubit states made up by Bell tensor product states constitute a faithful quantum channel and  the necessary and sufficient conditions a $2n$-qubit state has to fulfill for $n$-qubit deterministic teleportation are given in Ref.~\cite{PhysRevA.74.032324}.

Whereas the generation and the distribution of the quantum resource useful for 1-qubit teleportation, i.e., a single Bell state, has been widely investigated, the same does not hold for the entangled states used for $n$-qubit teleportation. In the context of spin chains, several schemes have been proposed to generate a Bell state between two distant 
qubits~\cite{PhysRevLett.91.207901,PhysRevLett.106.140501, PhysRevA.72.034303,PhysRevA.93.032310, PhysRevA.77.020303, PhysRevA.95.042335,PhysRevA.93.012343,depa2004,depa2005} 
based mainly on the same protocol used for one-qubit quantum state transfer. Clearly, any of these schemes could be used to sequentially generate Bell states by removing the entangled spins from the chain and wait that a new Bell pair is formed. A drawback of such a procedure is that it requires control over the motional degree of freedom of the spins and the sequential use of a spin chain as a quantum entangler could require its initialisation at each use, not to mention that the coherent dynamics of the quantum channel has to be preserved for longer times. It is evident hence that, also for scalability issues, it would be beneficial to have a {\textit{single}} quantum chain able to support the generation of $n$ pairs of Bell states, shared among a sender and a distant receiver, to be used as a resource for the teleportation of $n$ qubits.
Recently, the transfer of arbitrary two-qubit states, as well as specific classes thereof, 
have received a lot of attention~\cite{qst2,qst4,1367-2630-16-12-123003,qst5,PhysRevA.90.044301,0295-5075-119-3-30001,VIEIRA20182586,PhysRevA.92.022350}, but the search for a protocol able to generate, via the natural spin chain dynamics,  entanglement involving spins at distant locations to be used as a resource for 
two-qubit teleportation has yet not been addressed.   

In the present paper we address such a question. As far as a two-qubit  state is concerned, in \cite{PhysRevA.71.032303, PhysRevA.72.036301} it was shown that a perfect teleportation can be achieved by means of a four-qubit  maximally entangled state,  being the tensor product of a pair of two-qubit entangled states shared by the sender (A) and the receiver (B)
$\ket{\Psi}_{A_1 A_2 B_1 B_2}=\ket{\psi}_{A_1 B_1}\ket{\psi}_{A_2 B_2}$.
Here, we show how a 1D spin-$\frac{1}{2}$ chain with nearest-neighbor couplings of the $XX$-type and open boundary conditions can give rise to such a tensor product of Bell states of spins residing at the opposite edges of the chain, these being weakly coupled to the channel. The paper is organised as follows: in Sec.~\ref{S.Model} we introduce the spin model and in Sec.~\ref{S.Model2} we work out an effective perturbative Hamiltonian that faithfully reproduce its dynamics; in Sec.~\ref{S.Protocol} we evaluate the bipartite entanglement between the two pairs of spins located at the edges of the chain and we show its usefulness for a 2-qubit teleportation protocol; finally, in Sec.~\ref{S.Concl} we draw our conclusions and outlooks.

\section{The model}\label{S.Model}  
\begin{figure*}[t]
        \centering
          \includegraphics[width=0.9\linewidth]{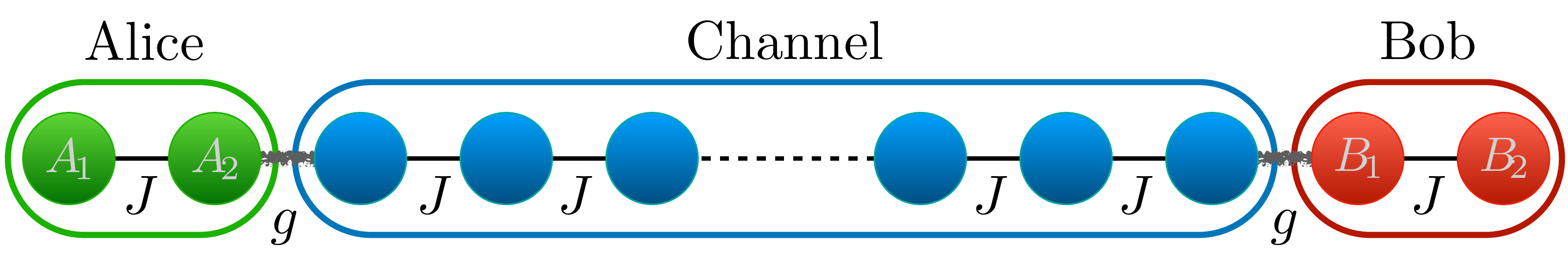}
        \caption{\label{n-QST}Alice (A) and Bob (B) each has access to a pair of qubits located at the opposite edges of a quantum channel. Their aim is to generate, via the natural dynamics of the spin chain, an entangled state $\ket{\Psi}_{A_1A_2B_1B_2}$ to be used as a resource for 2-qubit teleportation.} 
      \end{figure*}

Our model consists of a 1D spin-$\frac{1}{2}$ chain with open boundaries and isotropic nearest-neighbor interaction in the $XY$ directions, the first two spin are the sender (A) party, the last two are the receiver (B) party and the $M$ spins in between represent the channel:

\begin{equation}\label{E.XXHam}
H=\sum_{i}\frac{J_i}{2}\left(\hat{\sigma}_i^x\hat{\sigma}_{i+1}^x+\hat{\sigma}_i^y\hat{\sigma}_{i+1}^y\right),
\end{equation}
where $\hat{\sigma}^{\alpha}$ ($\alpha=x,y,z$) are the Pauli matrices and the index runs  through sender, channel and receiver: 
$i=A_1,A_2,1, \ldots ,M, B_1,B_2$. 
The Hamiltonian described by Eq.~\eqref{E.XXHam} exhibits $U(1)$ symmetry, conserving thus the 
total magnetisation in the $z$-direction, and reduces to a model of non-interacting spinless fermions, see Eq.~\eqref{E.Hfermrealspace}.
%, which will turn out to be useful in investigating the dynamics. 
The couplings $J$ are uniform along the chain but for the coupling of the pair of two-qubit blocks at the edges to the quantum channel, that is $J_i=J$ for $i\neq A_2, M$ and $J_i=g$ for $i= A_2, M$, with $g\ll J$ and $J = 1$ being the energy unit (see Fig.~\ref{n-QST}). A similar scheme has already be found to be successful for 2-qubit quantum state transfer in Ref.~\cite{qst5}.

%%%%%%%%%%%%%%%%%%%%%%%%%%%%%%%%%%%%%%%%%%%%%%%%%%%%%%%%%%%%%%%%%
%%%%%%%%%%%%%%%%%%%%%%%%%%%%%%%%%%%%%%%%%%%%%%%%%%%%%%%%%%%%%%%%%
Our protocol considers the case where the initial state of the quantum channel is fully polarized, and the pair of qubit in $A$ and in $B$ can be initialised, that is $\ket{\Phi}=\ket{\Psi}_A\ket{\bf{0}}\ket{\Psi}_B$, with $\ket{\bf{0}}=\ket{0_1 0_2 \dots 0_{M}}$. 
Given that we want to achieve a tensor product of two Bell states $\lbrace \ket{\Phi^{+}},\ket{\Phi^{-}},\ket{\Psi^{+}}, \ket{\Psi^{-}} \rbrace$ between pairs of spins at the opposite edges -- one instance of which would be, e.g., $\ket{\Psi}_{A_1,A_2,B_1,B_2}=\ket{\Phi^+}_{A_1,A_2}\otimes \ket{\Phi^+}_{B_1,B_2}$ --, we need to determine the Hamiltonian dynamics in the invariant subspaces with $\left(0,2,4\right)$, $\left(1,3\right)$ or $\left(2\right)$ flipped spins, depending of which Bell states enter the product. Nevertheless, as Bell states are equivalent under local unitary operations, each of the 16 tensor products can be obtained from an arbitrary one. Clearly, because of the conservation of the total magnetization along the $z$-axis, the initial number of flipped spin is conserved, therefore, in the $A \cup B$ block there have to be  $n\geq 2$ spin flipped. In the following we will investigate the case $n=2$.
The reason for such a choice is that, as we will show in our perturbation analysis, the quantum channel (the bulk of the chain) will not support any flipped spin during the dynamics. Therefore, in order to possibly generate a tensor product Bell state in the subspaces with $\left(0,2,4\right)$ or $\left(1,3\right)$, an initial state $\ket{\Phi}$ should be prepared which is not a tensor product of single-qubit states, implying that the initial state should contain some entanglement.  
Although this could be achieved exploiting the dynamics of the spins in block $A$ and $B$ before coupling them to the quantum channel, this would require an additional time-control over the couplings if the initial states are not eigenstates of their respective Hamiltonians.
 
Our analysis is thus made upon initial states of the form (using the notation $\ket{j_{A_1} j_{A_2} j_{B_1} j_{B_2}}=\ket{\underline{j_{A_1} j_{A_2}} 00\ldots 0 \underline{j_{B_1} j_{B_2}}}$):
\begin{align}
\ket{\Psi_{1}(0)} &= \ket{1100} \equiv \ket{\underline{11}00\ldots 0\underline{00}},\label{state1}\\
\ket{\Psi_{2}(0)} &= \ket{1010} \equiv \ket{\underline{10}00\ldots 0\underline{10}},\label{state2}\\
\ket{\Psi_{3}(0)} &= \ket{1001} \equiv \ket{\underline{10}00\ldots 0\underline{01}},\label{state3}\\
\ket{\Psi_{4}(0)} &= \ket{0110} \equiv \ket{\underline{01}00\ldots 0\underline{10}}.\label{state4} 
\end{align}
Note that the remaining options $\ket{0011}$ and $\ket{0101}$ 
are symmetric to $\ket{\Psi_{1}(0)}$ and $\ket{\Psi_{2}(0)}$, respectively.

Eq.~\eqref{E.XXHam} can be mapped to a spinless fermion model via the Jordan-Wigner transformation~\cite{LIEB1961407}
\begin{equation}
\label{E.Hfermrealspace}
H=\sum_{i=1}^{N-1}J_i \left(\hat{c}^{\dagger}_i \hat{c}_{i+1} {+}\hat{c}_i \hat{c}^{\dagger}_{i+1}\right),
\end{equation}
where and $N=M+4$, $\hat{c}_{i}^{\dagger}$ and $\hat{c}_{i}$ are, respectively, fermionic creation and annihilation
operators at site $i$.
Because of the quadratic nature of the Hamiltonian, the one-particle spectrum is sufficient to describe the full
dynamics in every $n$-flipped spin sector. Denoting by $\varepsilon_k$ and $\ket{\varepsilon_k}=\hat{c}_k^\dagger \ket{\{0\}}$
the single-particle $k$-th energy eigenvalue and its corresponding eigenvector,
the full Hamiltonian operator acting on a $2^N$ dimensional
Hilbert space, is easily decomposed into a direct sum over all
particle number-conserving invariant subspaces
$H=\bigoplus_{n=1}^N H_n$, where
\begin{widetext}
\begin{equation}\label{E.decom}
H_n{=}\!\!\!\!\!\!\!\!\sum_{k_1{<}k_2{<}...{<}k_n{=}1}^N
\!\!\!\!\!\!\!\!\!\!\!\!
\left(\varepsilon_{k_1}{+}\varepsilon_{k_2}{+}...{+}\varepsilon_{k_n}\right)
\hat{c}^{\dagger}_{k_1}\hat{c}^{\dagger}_{k_2}...\hat{c}^{\dagger}_{k_n}\ket{\{0\}}\bra{\{0\}}\hat{c}_{k_n}...\hat{c}_{k_2}\hat{c}_{k_1}.
\end{equation}
\end{widetext}
Each $H_n$, therefore
can be constructed quite simply once the single-particle spectrum
is known. Notice that the specific ordering of the $k_i$\rq{}s in
the sum of Eq.~\eqref{E.decom} is taken in such a way that
unwanted phase factors do not arise when mapping back into spin
operators via the inverse Jordan-Wigner transformation. 
Every invariant subspace is spanned by a set of states having a fixed number of flipped spins. Hence, one can define 
a single-particle states obtained by flipping the $j$-th spin of the system $\ket{j}=\hat{c}_j^\dagger \ket{\{0\}}$, the two-particle states, obtained flipping the $j$-th and $i$-th sites of the system (with $j<i$)  $\ket{ji}=\hat{c_j}^\dagger \hat{c_i}^\dagger \ket{\{0\}}$, and so on. 
The non-interacting nature of the fermionic Hamiltonian in Eq.~\eqref{E.decom} allows to reduce the two-particle transition 
amplitudes $h_{nm}^{pq}(t)=\bra{pq}e^{-i t H_2}\ket{nm}$ to determinants of matrices whose elements 
are single-particle transition amplitudes $f_i^j(t)=\bra{j}e^{-i t H_1}\ket{i}$, 
where $i{=}\{n,m\}$ and $j{=}\{p,q\}$ (see, e.g., Ref.~\cite{qst4,1367-2630-16-12-123003}):
\begin{equation}\label{E.2to1}
h_{nm}^{pq}(t){=}  \begin{vmatrix}

        f_n^p(t) & f_n^q(t) \\

 	f_m^p(t) & f_m^q(t) \\

    \end{vmatrix}~.
\end{equation} 
Consequently, the evolved state in the two-particle sector results
\begin{equation}\label{E.2state}
\ket{\Psi(t)}=\sum_{p<q=1}^N h_{n_{0}m_{0}}^{pq}(t)\ket{pq}~,
\end{equation}
when starting from the initial state $\ket{n_{0}m_{0}}$. 

Notwithstanding we are able to solve the exact full dynamics of the model in Eq.~\eqref{E.XXHam} numerically, it is instructive to rely on a perturbative approach, due to the presence of the weak couplings $g$, in order to derive an effective Hamiltonian allowing us to identify more easily the peculiar dynamical behaviour behind the generation of highly entangled states between blocks A and B. 

\section{Perturbative analysis}\label{S.Model2}

Similarly to the one-particle subspace dynamics, the model supports one- and two-particle Rabi-like oscillations between its edge spins. A similar argument has been used to generate a single Bell state in the one-particle subspace~\cite{PhysRevA.72.034303} in a $N$-spin chain with a single weakly coupled spin residing at each end, given at half the QST time one has $\ket{1_{A_1}0_N} \rightarrow\frac{1}{\sqrt{2}}\left(\ket{1_10_N}-\ket{0_11_N}\right)$. 

However a straightforward extension with two non-interacting edge spins, such as proposed in Refs.~\cite{VIEIRA20182586}, each weakly coupled to the edge spins of the quantum channel, does not yield a tensor product of Bell states. Indeed, starting from, e.g., $\ket{1_{A_1}1_{A_2}}$, because of permutation symmetry of the edge spins, the amplitudes of the states $\ket{1_{A_i}1_{B_j}}$ ($i,j=1,2$)
have to be equal at all times, preventing thus the generation of a Bell state between any pair of the sender-receiver block.    

In the $n=2$ flipped spin subspace
the states whose dynamics we are about to investigate are listed in Eqs. \eqref{state1}-\eqref{state4}.
Given the dynamics is restricted to the two-excitation subspace and that 
$g\ll J$ our task now is to carry out a perturbative approach in order to derive an effective Hamiltonian
involving only the six possible configurations spanning over both edge blocks ($A$ and $B$), that is
$\lbrace \ket{A_1B_1},\ket{A_1B_2},\ket{A_2B_1},\ket{A_2B_2},\ket{A_1A_2},\ket{B_1B_2} \rbrace$.

\subsection{Effective description}

Let us split Hamiltonian \eqref{E.Hfermrealspace} into $H = H_{0}+H_{\mathrm{ch}}+H_{I}$, where
\begin{align}
H_{0} &= J(c_{A_{1}}^{\dagger}c_{A_{2}}+c_{B_{1}}^{\dagger}c_{B_{2}} + \mathrm{H.c.}),\\
H_{\mathrm{ch}} &=\sum_{i=1}^{M-1} J(c_{i}^{\dagger}c_{i+1}+ \mathrm{H.c.}),\\
H_{I} &= g(c_{A_{2}}^{\dagger}c_{1}+c_{M}^{\dagger}c_{B_{1}} + \mathrm{H.c.}).
\end{align}
The effective Hamiltonian can be obtained via a second-order perturbation method that gives \cite{li05}
\begin{widetext}
\begin{equation}\label{Heff}
\bra{\psi_{0,i}} H_{\mathrm{eff}} \ket{\psi_{0,j}} =  E_{j}\delta_{i,j} - \frac{1}{2}\sum_{k}\left[ \frac{(H_{I})_{ik}(H_{I})_{kj}}{\lambda_{k}-E_{i}} +  \frac{(H_{I})_{ik}(H_{I})_{kj}}{\lambda_{k}-E_{j}} \right],
\end{equation}
\end{widetext}
where $(H_{I})_{ik}\equiv \bra{\psi_{0,i}} H_{I} \ket{\lambda_{k}}$ and 
$\lbrace \ket{\psi_{0,i}} \rbrace$ and $\lbrace\ket{\lambda_{k}}\rbrace$ are, respectively, the eigenstates of $H_{0}$ and $H_{\mathrm{ch}}$, with corresponding energies $\{E_{i}\}$ and $\{\lambda_{k}\}$.

The unperturbed eigenstates of the subsystem of interest, i.e., 
blocks A and B read
\begin{align}
\ket{\psi_{0,1}} &=  \ket{A_{+}B_{+}},\\ 
\ket{\psi_{0,2}} &=   \ket{A_{+}B_{-}},\\ 
\ket{\psi_{0,3}} &=  \ket{A_{-}B_{+}}, \\ 
\ket{\psi_{0,4}} &=  \ket{A_{-}B_{-}}, \\ 
\ket{\psi_{0,5}} &=  \ket{A_{1}A_{2}},\\ 
\ket{\psi_{0,6}} &=  \ket{B_{1}B_{2}},
\end{align}
where
$\ket{A_{\mu}B_{\nu}} = (\ket{A_{1}}+\mu \ket{A_{2}})\otimes (\ket{B_{1}}+\nu \ket{B_{2}})/2,$
with $\mu, \nu = \pm$.
Their corresponding eigenvalues are $E_{1} = 2J, E_{4} = -2J$, and $E_{2,3,5,6} = 0$. 

The single-particle eigenstates of the channel ($H_{\mathrm{ch}}$)  are
\begin{equation}\label{eigenV}
\ket{\epsilon_{m}} = \sqrt{\dfrac{2}{M+1}}\sum_{x=1}^{M}\mathrm{sin}\left( \dfrac{\pi mx}{M+1}\right)\ket{x},
\end{equation}
with energies $\epsilon_m= 2J\mathrm{cos}(\frac{\pi m}{M+1})$. So one can construct $4\times M $ unperturbed states as  $\ket{\lambda_{k=(l,m)}} = \ket{\eta_{l}}\ket{\epsilon_{m}}$ ($l=1,2,3,4$ and $m=1,\ldots,M$), with $\ket{\eta_{1,2}}  = (\ket{A_{1}}\pm\ket{A_{2}})/\sqrt{2}$ and $\ket{\eta_{3,4}}  = (\ket{B_{1}}\pm\ket{B_{2}})/\sqrt{2}$. 
The corresponding eigenvalues read $\lambda_{l,m} = \epsilon_{m}+J$ for $l=1,3$ and $\lambda_{l,m} = \epsilon_{m}-J$ for $l=2,4$ 
The remaining unperturbed eigenstates 
involve linear combinations of states containing no excitations in either block $A$ or $B$. Those provide no contribution to the sum in Eq. \eqref{Heff} given $\langle \psi_{0,i} \vert H_{I} \vert x_{1}x_{2} \rangle = 0$ for all $i$ and $x_1,x_2 \in \{ 1,\ldots,M \}$. 

With all the above relations at hand, one is able to evaluate the matrix elements of $H_{\mathrm{eff}}$
through Eq. \eqref{Heff}. When doing so, one will find expressions such as
\begin{equation} 
\Lambda_{1}^{\pm}  =\frac{g^2}{2}\sum_{m} \frac{a_{m}^{2}}{\epsilon_{m}\pm J},\,\,\,\,\,\, 
\Lambda_{2}^{\pm} = \frac{g^2}{2}\sum_{m} \frac{a_{m}^{2}e^{2i\theta_{m}}}{\epsilon_{m}\pm J},
\end{equation}
where we are considering a mirror-symmetric channel fulfilling
$|\alpha_{1}^{m}| =|\alpha_{M}^{m}| = a_{m}$ and
$(\alpha_{1}^{m}) = (\alpha_{M}^{m})^{*} = a_{m}e^{i\theta_{m}}$, with $\alpha_{x}^{m} \equiv \langle x \vert \epsilon_{m} \rangle$~\cite{doi:10.1063/1.4797477}. These result in $\Lambda_{1}^{\pm} = 0$ and $\Lambda_{2}^{\pm}=-g^{2}/2J$ if $M = 6n$ ($n = 1,2,\ldots$) and $\Lambda_{1}^{\pm} = \pm g^{2}/2J$ and $\Lambda_{2}^{\pm}=g^{2}/2J$ for $M=6n+4$. We also point out that the above perturbation approach is not valid for $M=6n+2$ given it yields $\epsilon_{m} = \pm J$ thereby causing divergence of the sums above. Without loss of generality, though, we consider $M=6n$ for the remainder of this paper. 

After working out every term of the effective Hamiltonian via Eq. \eqref{Heff}, its matrix form written in the basis $\{ \ket{\psi_{0,i}} \}$ reads 
\begin{equation} \label{Heff2}
H_{\mathrm{eff}}=\left(
\begin{array}{cccccc}
 2 J & 0 & 0 & 0 & \frac{g^2}{2J} & \frac{g^2}{2J} \\
 0 & 0 & 0 & 0 & \frac{g^2}{2J} & -\frac{g^2}{2J} \\
 0 & 0 & 0 & 0 & \frac{g^2}{2J} & -\frac{g^2}{2J} \\
 0 & 0 & 0 & -2 J & \frac{g^2}{2J} & \frac{g^2}{2J} \\
 \frac{g^2}{2J} & \frac{g^2}{2J} & \frac{g^2}{2J} & \frac{g^2}{2J} &
   0 & 0 \\
 \frac{g^2}{2J} & -\frac{g^2}{2J} & -\frac{g^2}{2J} & \frac{g^2}{2J}
   & 0 & 0 \\
\end{array}
\right).
\end{equation}
We finally express its eigenvectors and corresponding eigenvalues in terms of $\{ \ket{j_{A_{1}}j_{A_{2}}j_{B_{1}}j_{B_{2}}} \}$, with $j_{x}\in\{ 0,1 \}$ (e.g., $\ket{A_{1}B_{1}} \equiv \ket{1010}$):
\begin{align}
\ket{\xi_{1}} & {\approx} \frac{1}{2}(\ket{1010}{+}\ket{1001}{+}\ket{0110}{+}\ket{0101}), \label{e1}\\
\ket{\xi_{2}} & {\approx} \frac{1}{2}(\ket{0101}{-}\ket{1001}{-}\ket{0110}{+}\ket{0101}), \label{e2}\\
\ket{\xi_{3}} & {\approx} \frac{1}{\sqrt{2}}(\ket{1100}{+}\ket{0011}), \label{e3} \\
\ket{\xi_{4}} & {=} \frac{1}{\sqrt{2}}({-}\ket{1001}{+}\ket{0110}),  \label{e4}\\
\ket{\xi_{5}} & {=} \frac{1}{2}(\ket{1010}{-}\ket{0101}{+}\ket{1100}{-}\ket{0011}), \label{e5}\\
\ket{\xi_{6}} & {=}\frac{1}{2}(\ket{1010}{-}\ket{0101}{-}\ket{1100}{+}\ket{0011}), \label{e6}
\end{align}
with $\xi_{1}{ \approx}  2J$, $\xi_{2}{\approx}  {-}2J$, $\xi_{3}{ =} \xi_{4} {=} 0$, $\xi_{5} {=} g^{2}/J$, and $\xi_{6}{=} {-}g^{2}/J$.  

\begin{figure}[t]
 \centering
    \includegraphics[width=0.5\textwidth]{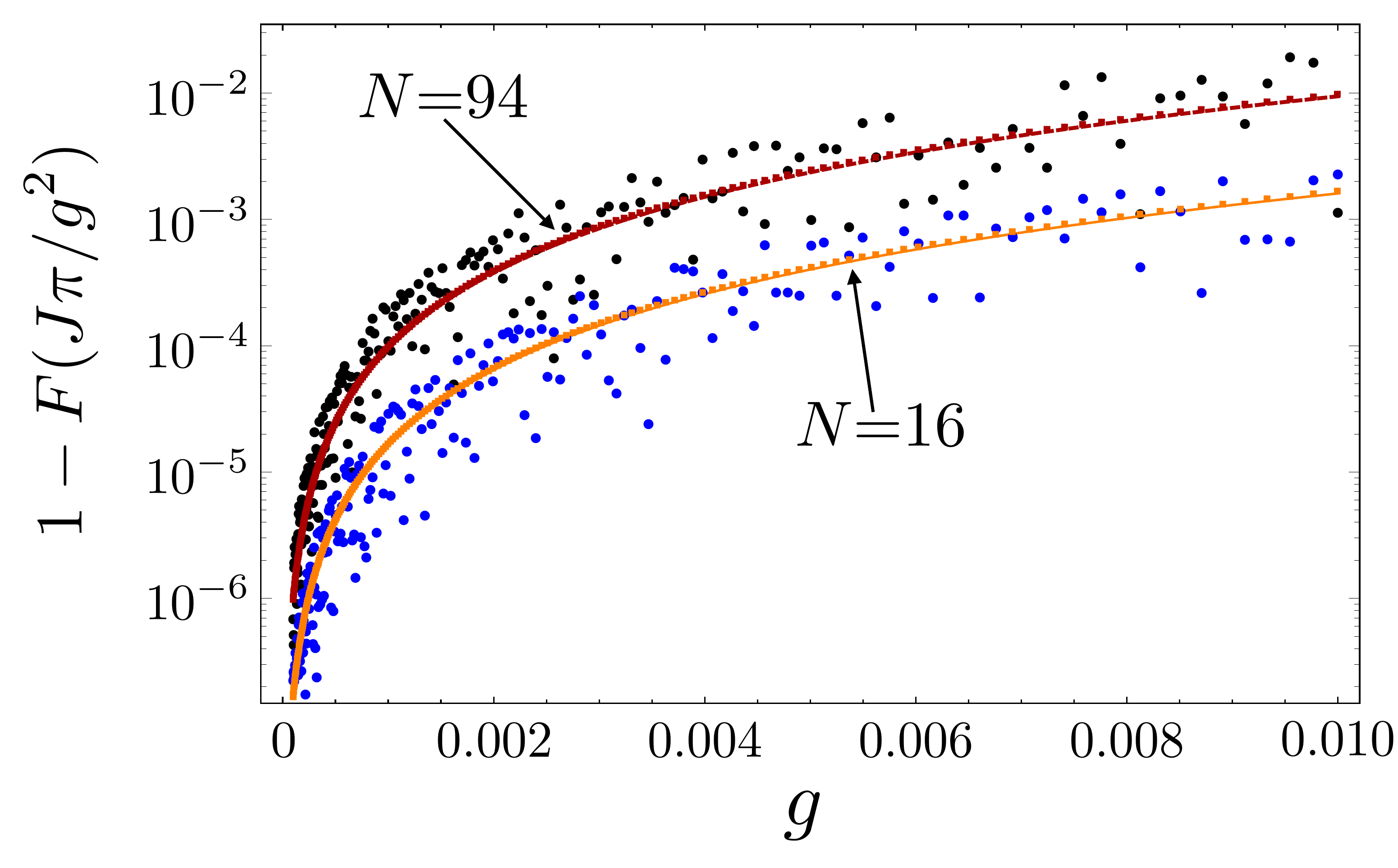}
    \caption{Infidelity between exact dynamics and perturbative one versus the coupling $g$ for $N=94$ and $N=16$ evaluated at $t=\pi J/g^2$. We take as initial state $\ket{\Psi_1(0)}$.}
     \label{InFID}
\end{figure}

In Fig.~\ref{InFID} we report the infidelity between the states obtained via Eq.~\ref{E.Hfermrealspace} and Eq.~\ref{Heff2} for the initial state $\ket{\Psi_1(0)}$ and for different values of $g$ and $N$. We see that the infidelity scales as $Ng^2$, validating thus the second-order perturbation approach for $g\ll\frac{1}{\sqrt{N}}$

\subsection{Generation of Bell product states}
We are now ready to track down the time evolution of the initial states displayed in Eqs. \eqref{state1}-\eqref{state4}
in the light of second-order perturbation theory and check whether a tensor product of Bell states can be achieved involving blocks $A$ and $B$. We stress that the effective description in Eq. \eqref{Heff2} entails
no excitation within the channel at any time. 

\begin{figure}[t]
 \centering
    \includegraphics[width=0.5\textwidth]{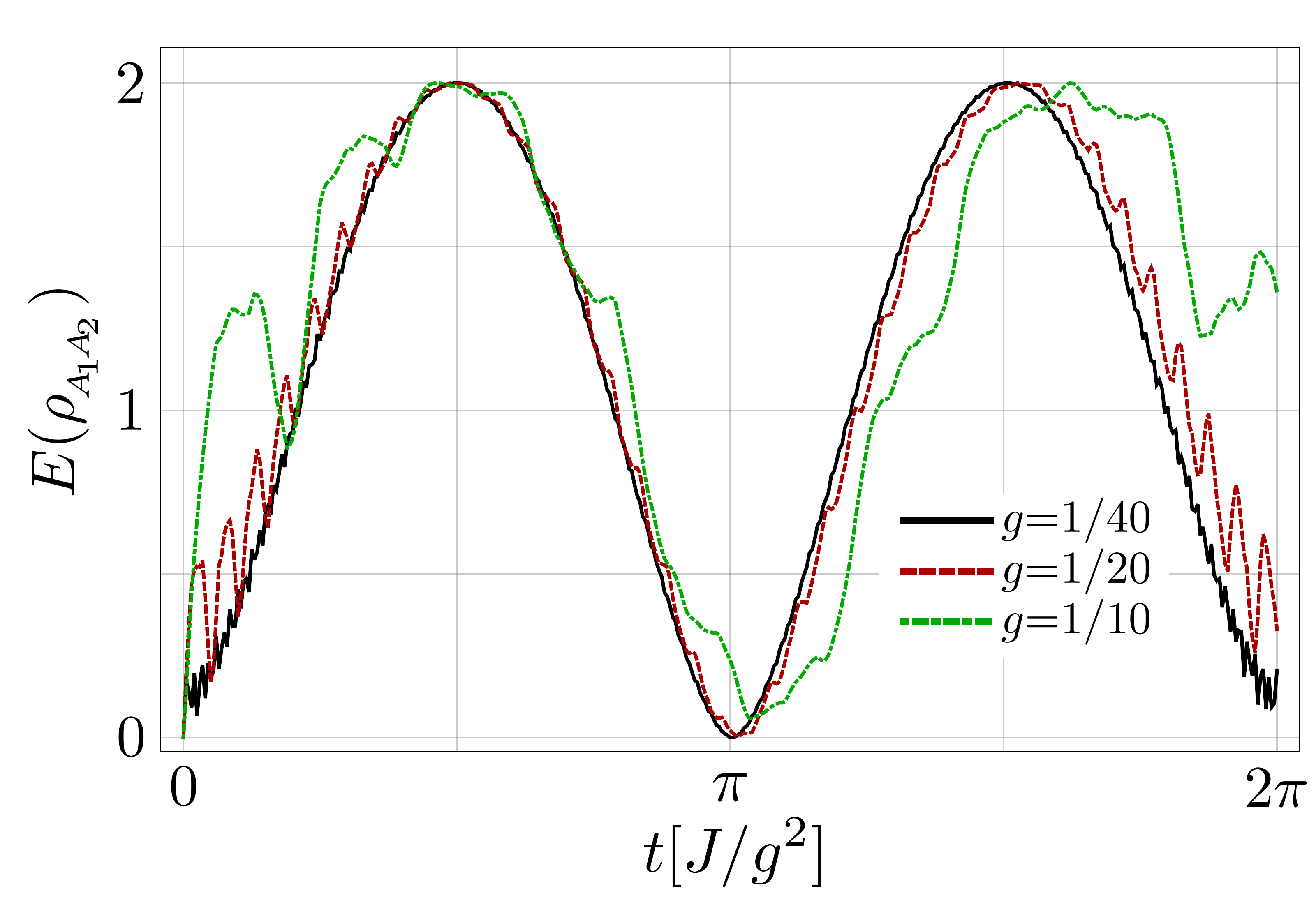}
    \caption{Time evolution of the entanglement entropy  $E(\rho_{A_1A_2})$ for $\ket{\Psi_{1}(0)} = \ket{1100}$ and different values of $g$. Maximum entanglement is achieved at times $t^*=\frac{(2n+1)\pi J}{2 g^2}$, with $n$ being a positive integer, when the state can be expressed as a tensor product of Bell states shared by pairs $(A_1, B_2)$ and $(A_2,B_1)$.}
     \label{EE.1}
\end{figure}

According to the eigenstates obtained above, we arrive at the following dynamics for $\ket{\Psi_{1}(0)} = \ket{1100}$:
\begin{align}
\ket{\Psi_1(t)}&=\frac{1}{2}\left[\left(1-\cos \tfrac{g^2 t}{J}\right)\ket{0011} +i \sin\tfrac{g^2 t}{J} \ket{0101} \right. \nonumber \\
&\,\,\,\, \,\,\left. -i \sin \tfrac{g^2 t}{J}\ket{1010}+ \left(1+\cos \tfrac{g^2 t}{J}\right)\ket{1100}\right]. \label{E.D1}
\end{align}
Given the above is a pure state, we can evaluate the amount of entanglement block A is sharing with block B by means of the entanglement entropy $E(\rho_{A_1A_2}) = Tr[\rho_{A_1A_2}\log_{2}\rho_{A_1A_2}]$, with $\rho_{A_1A_2}(t) = Tr_{B_1B_2}(\ket{\Psi_{1}(t)}\bra{\Psi_{1}(t)})$, which is reported in Fig.~\ref{EE.1} wherein
we check it reaches the maximum value attainable for two qubits, $E=2$, at $t^*=\frac{(2n+1)\pi J}{2 g^2}$, with $n=0,1,2,\ldots$. At such times, the state of Eq.~\eqref{E.D1} reads
\begin{align}
\ket{\Psi_1\left(t^*\right)}&=\frac{1}{2}\left(\ket{0011} +(-1)^n i \ket{0101} \right. \nonumber \\
& \,\,\,\,\,\,\left. +  (-1)^{n+1} i  \ket{1010}+ \ket{1100}\right).\label{E.dyn1_t}
\end{align}
This state can be also written as a tensor product state of two Bell states between the pairs $(A_1,B_2)$ and $(A_2,B_1)$, namely
$\ket{\Psi_1\left(t^*\right)}=\ket{\phi}_{A_1B_2}\otimes \ket{\phi}_{A_2B_1}$, where
\begin{align}\label{E.dyn1_t_Bell}
\ket{\phi}_{A_1B_2}&=\frac{1}{\sqrt{2}}\left(\ket{01} +(-1)^{n+1} i \ket{10} \right), \\
\ket{\phi}_{A_2B_1}&=\frac{1}{\sqrt{2}}\left(\ket{01} +(-1)^n i \ket{10} \right)~.
\end{align}

Although the state in Eq.~\eqref{E.dyn1_t} is a legitimate one for two-qubit teleportation, Alice may apply a single-qubit phase gate $R\left(\frac{\pi}{2}\right)$ to retrieve the standard Bell states $\ket{\Psi^{\pm}}$ and subsequently follow the protocol addressed in Ref.~\cite{PhysRevA.71.032303} to carry out the teleportation. Otherwise, as pointed out in Ref.~\cite{PhysRevLett.70.1895}, there will be a different set of two local unitary operations Bob has to perform on each of his qubits which we report in the following section.

We reach to a similar scenario starting from $\ket{\Psi_2(0)} = \ket{1010}$,
\begin{align}
\ket{\Psi_2(t)}&=\frac{1}{2}\left[\left(\cos 2Jt+\cos \tfrac{g^2t}{J}\right)\ket{1010} \right.\nonumber \\
&\,\,\,\,\,\, \left. +\left(\cos 2Jt-\cos \tfrac{g^2t}{J}\right)\ket{0101}\right. \nonumber \\
&\,\,\,\,\,\,\left. -i \sin 2Jt\left(\ket{1001}+\ket{0110}\right)\right. \nonumber \\
&\,\,\,\,\,\,\left. -i \sin \tfrac{g^2t}{J} \left(\ket{1100}-\ket{0011}\right)\right],
\label{E.D2}
\end{align}
with maximum entanglement entropy $E(\rho_{A_1A_2})$ at the same time $t^*=\frac{(2n+1)\pi J}{2 g^2}$ when the state reads
\begin{align}
\ket{\Psi_2\left(t^*\right)}&=\frac{1}{2}\left[i(-1)^n\left(\ket{0011}-\ket{1100}\right)\right.\nonumber \\
&\,\,\,\,\,\, \left. +\cos 2Jt^*\left(\ket{0101}+\ket{1010}\right)\right. \nonumber \\
&\,\,\,\,\,\,\left. -i\sin 2Jt^*\left(\ket{0110}+\ket{1001}\right)\right].
\label{E.dyn2_t}
\end{align}
If we now assume that the ratio $J^2/g^2$ is commensurate and such that $2Jt^*{=}2m\pi$ or $2Jt^*{=}(2m{+}1)\pi$, we have that the $\cos$ and $\sin$ functions are, respectively, $\pm 1$ and 0.
The state in Eq.~\eqref{E.dyn2_t} becomes
\begin{align}
\ket{\Psi_2\left(t^*\right)}&=\frac{1}{2}\left[i(-1)^n\left(\ket{0011}-\ket{1100}\right)\right. \nonumber \\
& \,\,\,\,\,\, \left.+\mu_{n}\left(\ket{0101}+\ket{1010}\right)\right],
\label{E.dyn21_t}
\end{align} 
with  
$\mu_{n} \equiv \mathrm{Sign}[\cos 2Jt^*]$, 
which can be readily seen to be the product state $\ket{\Psi_2\left(t^*\right)}=\ket{\phi}_{A_1B_2}\otimes \ket{\phi}_{A_2B_1}$, where
\begin{align}\label{E.dyn21_t_Bell}
\ket{\phi}_{A_1B_2}&=\frac{1}{\sqrt{2}}\left(\ket{01}-\mu_{n}(-1)^{n} i \ket{10} \right), \\
\ket{\phi}_{A_2B_1}&=\frac{1}{\sqrt{2}}\left(i (-1)^{n} \ket{01} +\mu_{n} \ket{10} \right).
\end{align}
Similarly, for $2Jt^*=\pi/2+2n \pi$ or $2Jt^*=3\pi/2+2n \pi$,  $\cos$ and $\sin$ functions are, respectively, 0 and $\pm 1$, and the state in Eq~(\ref{E.dyn2_t}) evolves into the Bell product state  $\ket{\Psi_2\left(t^*\right)}=\ket{\phi}_{A_1B_1}\otimes \ket{\phi}_{A_2B_2}$. For times different from those reported above, altough the entanglement entropy is maximum, the state can not be decomposed into a tensor product of Bell states.

The two remaining initial states in our investigation, $\ket{\Psi_{3}(0)} = \ket{1001}$ and $\ket{\Psi_{4}(0)} =\ket{0110}$, do not yield any entanglement between block A and B at any time, that is $E(\rho_{A_1A_2}(t)) = 0$. It is interesting to note that their dynamics does not even involve $g$ in second-order perturbation expansion.

\section{Entanglement of teleportation}\label{S.Protocol}
The 1-qubit teleportation protocol establishes that A and B share a pair of qubits in a maximally entangled state $\ket{\Psi^k_\theta}$ and that the sender (A) performs a Bell-measurement 
on its shared qubit and an unknown one. The result of such a measurement, encoded in two classical bits has to be sent to the receiver (B) in order to put him in the condition to choose the right operation on its shared qubit  leaving him with the unknown teleported state $\ket{\varphi}$. 
This means that  the initial state of the protocol can be written as
\begin{align}\label{1Q-A_ops}
\ket{\phi}{\otimes}\ket{\Psi^k_\theta}=\frac{1}{2}\sum_{j=1}^4\ket{\Psi^j_\theta}{\otimes}\mathcal{O}^k_j \ket{\varphi}
\end{align}
where $\ket{\Psi^j_\theta}$ stand for generalized Bell-states (note that the entanglement of a Bell state is independent of the relative phase) 
\begin{align}
\ket{\Psi^{1,2}_{\theta}}=\frac{1}{\sqrt{2}}\left(\ket{01}\mp e^{-i \theta}\ket{10}\right)\\
\ket{\Psi^{3,4}_{\theta}}=\frac{1}{\sqrt{2}}\left(\ket{00}\mp e^{-i \theta}\ket{11}\right)
\end{align}
and the operators $\mathcal{O}^k_j$ depend on the shared entangled state.
Alice now perform a Bell-measurement depending on the relative phase $\theta$, obtaining with equal probability one of $\ket{\Psi^j_\theta}$ states and classically comunicate her result.
At this point Bob is able to recover the unknown state $\ket{\varphi}$ performing the right operation $\mathcal{\tilde O}_j^k$ (according to initial shared state) among the set
\begin{align}
\mathcal{\tilde O}_j^1=& \lbrace -R(\theta) ,R(\theta)\sigma^z,\sigma^x, \sigma^z\sigma^x \rbrace\\
\mathcal{\tilde O}_j^2=& \lbrace -R(\theta)\sigma^z,R(\theta) , \sigma^z\sigma^x,\sigma^x \rbrace\\
\mathcal{\tilde O}_j^3=& \lbrace -R(-\theta)\sigma^x,-R(-\theta)\sigma^z\sigma^x, \mathbf{I} ,-\sigma^z \rbrace\\
\mathcal{\tilde O}_j^4=& \lbrace R(-\theta)\sigma^z\sigma^x,R(-\theta)\sigma^x,-\sigma^z ,\mathbf{I} \rbrace\\
\end{align}
with 
\begin{align}R(\theta)=\begin{pmatrix}e^{i\theta}&0\\0& e^{-i\theta}\end{pmatrix}\end{align}
observe that $\mathcal{\tilde O}_j^k$ are simply the inverse of $\mathcal{O}_j^k$ in \eqref{1Q-A_ops}.
The quantum resource for this protocol is a maximally entangled state, i.e., the Bell state of two qubits. 

For 2-qubit teleportation protocol let us suppose Alice and Bob share the state 
%\ket{\Psi^{k_1k_2}_\theta}{\equiv}
$\ket{\Psi^{k_1}_\theta}_{A_1B_2}{\otimes}\ket{\Psi^{k_2}_\theta}_{A_2B_1}$ 
and Alice want to teleport an arbitrary two-qubit state $\ket{\varphi}_{XY}{=} a\ket{00}{+}b\ket{01}{+}c\ket{10}{+}d\ket{11}$. 
The initial state of the protocol can be so written as
\begin{align}
%\ket{\varphi}_{\!XY}{\otimes}\ket{\Psi^{k_1k_2}_\theta}=
&\ket{\varphi}_{\!X\!Y}{\otimes}\ket{\Psi^{k_1}_\theta}_{\!\!A_1B_2}{\otimes}\ket{\Psi^{k_2}_\theta}_{\!\!A_2B_1}{=} \\
&\frac{1}{4}\sum_{j_1,j_2=1}^4\ket{\Psi^{j_1}_\theta}_{\!\!X\!A_1}\!\!{\otimes}\ket{\Psi^{j_2}_\theta}_{\!\!Y\!A_2}\!{\otimes}\left(\mathcal{O}^{k_2}_{j_1}{\otimes}
\mathcal{O}^{k_2}_{j_2}\right)\ket{\varphi}_{\!B_1\!B_2}\nonumber
\end{align}
As a consequence, a measurement in the generalised Bell basis given above on Alice's pairs of qubits $(X,A_1)$ and $(Y,A_2)$, 
reduces the scheme to the standard single-qubit teleportation protocol for each qubit $X$ and $Y$. 

In Ref.~\cite{PhysRevA.71.032303}, $E_T\left(\ket{\Psi}\right)$, the Entanglement of Teleportation (EoT),  has been introduced as a measure of the usefulness of a $2n$-qubit pure state, $\ket{\Psi}$, for $n$-qubit teleportation. In the following we report its expression for the case of four qubits. The EoT is based on the generalized concurrence~\cite{PhysRevA.63.044301}, $C\left(\ket{\Psi}\right)=|\bra{\Psi}{\tilde{\Psi}}\rangle|$, where $\ket{{\tilde{\Psi}}}={\hat{\sigma}}^y_{A_{1}}{\hat{\sigma}}^y_{A_{2}}{\hat{\sigma}}^y_{B_{1}}{\hat{\sigma}}^y_{B_{2}}\ket{\Psi}^*$ and the state is expressed in the computational basis. Hence, $E_T\left(\ket{\Psi}\right)=\frac{1}{16}\sum_{i=1}^{16}C\left(\ket{\Psi^{(i)}} \right)$, where $\ket{\Psi^{(i)}}$ are all the orthogonal states that can be obtained from $\ket{\Psi}$ by applying certain single-qubit unitary operations, as reported in Ref.~\cite{PhysRevA.71.032303}.
Let us point out that the EoT is independent of the choiche of basis as long as each of the 16 basis states are composed of tensor product of maximally entangled states. 
Straightforward calculations shows that in the case of $\theta=\frac{\pi}{2}$ the states reported in Eqs.~\eqref{E.dyn1_t} and \eqref{E.dyn21_t} have unit EoT.

Although the states obtained by the full and reduced dynamics have vanishing infidelity, as shown in Fig.~\ref{InFID}, let us also compare, for the sake of completeness, the efficiency of the teleportation protocol performed via the exact and the reduced states, as, in principle, states with high fidelity may not share the same reources~\cite{PhysRevA.89.012305}. 
To this aim we report the Fidelity of Teleportation, $F_T$, by means of the full and the effective Hamiltonians reported in the previous sections, in Eqs. \eqref{E.Hfermrealspace} and \eqref{Heff2}, respectively. The Fidelity of Teleportation is given by the overlap of the unknown state to be teleported, say $\ket{\varphi_{in}}$ and the Bob's output state, ${\hat{\rho}}_{out}$, $F_T=\bra{\varphi_{in}}{\hat{\rho}}_{out}\ket{\varphi_{in}}$. Using the two-qubit parametrization as in Ref.~\cite{qst4}
\begin{widetext}
\begin{align}
\ket{\varphi}_{XY}=&
\sqrt{\frac{1-s}{2}}\left(\cos\frac{\theta_1}{2}\ket{0}+e^{i\phi_1}\sin\frac{\theta_1}{2}\ket{1}\right)\otimes\left(\cos\frac{\theta_2}{2}\ket{0}+e^{i\phi_2}\sin\frac{\theta_2}{2}\ket{1}\right)+ \nonumber \\
&\sqrt{\frac{1+s}{2}}\left(e^{-i\phi_1}\sin\frac{\theta_1}{2}\ket{0}-\cos\frac{\theta_1}{2}\ket{1}\right)\otimes\left(e^{-i\phi_2}\sin\frac{\theta_2}{2}\ket{0}-\cos\frac{\theta_2}{2}\ket{1}\right)
\end{align}
\end{widetext}
with $0\leq\theta_{1,2}\leq\pi$ , $0\leq\phi_{1,2}\leq2\pi$ , and $ -1\leq s\leq 1$.

After working out the fidelity of teleportation of such a state, according to the effective Hamiltonian description,
we integrate it over all possible inputs to obtain the average fidelity of teleportation
\begin{align}
\bar{F}_{\text{eff}}(t)=\frac{1}{2}-\frac{7}{54}\cos \frac{2 g^2t}{J}+\frac{10}{27}\sin \frac{g^2t}{J}~.
\end{align}

\begin{figure}[t]
 \centering
    \includegraphics[width=0.5\textwidth]{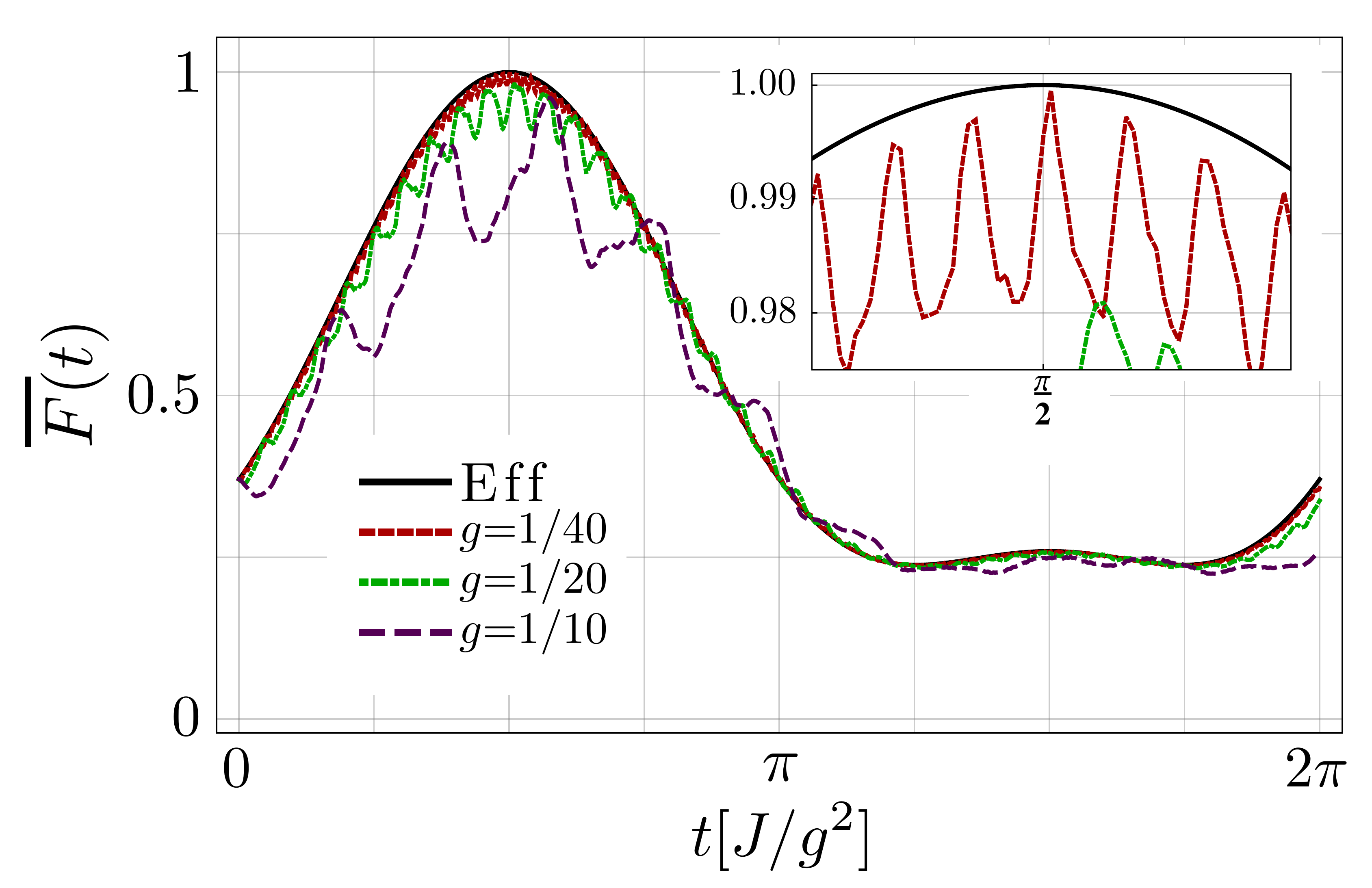}
    \caption{Average fidelity of teleportation for the full model for a random two-qubit state and $N=22$, $g=0.1J$ (red), $g=0.05J$ (green), and $g=0.025J$ (purple). The blue curve represents the same quantity for the effective dynamics obtained from the second-order perturbation approach.}
     \label{FMean}
\end{figure}

On the other hand, the average $F_T$ for the full dynamics reads
\begin{widetext}
\begin{align}
\bar{F}_{T}(t)&=\frac{1}{27}\left(7+3 \vert h_{{}_{12}}\vert^2+3\vert h_{{}_{1N{-}1}}\vert^2+6\vert h_{{}_{1N}}\vert^2+3\vert h_{{}_{N{-}1N}}\vert^2+3\vert h_{{}_{2N}}\vert^2 \right. \nonumber \\
&\,\,\,\,\,\,\left. -2\sum_{n=3}^{N-2}\left(\vert h_{{}_{1n}}\vert^2+\vert h_{{}_{2n}}\vert^2+\vert h_{{}_{nN{-}1}}\vert^2+\vert h_{{}_{nN}}\vert^2\right)+
14 Re (h_{{}_{12}}h^*_{{}_{N{-}1N}}-h_{{}_{2N}}h^*_{{}_{1N{-}1}})\right.\nonumber\\
&\,\,\,\,\,\,\left. +10 Im(h_{{}_{12}}h^*_{{}_{1N{-}1}}+h_{{}_{2N}}h^*_{{}_{N{-}1N}}-h_{{}_{1N{-}1}}h^*_{{}_{N{-}1N}}-h_{{}_{12}}h^*_{{}_{2N}})-
4\sum_{n=3}^{N-2}Im(h_{{}_{1n}}h^*_{{}_{nN}}-h_{{}_{2n}}h^*_{{}_{nN{-}1}})
\right),
\end{align}
\end{widetext}
where for clarity $h_{{}_{pq}}$ stands for $h^{pq}_{12}$ as defined in Eq.~\eqref{E.2to1}, and $Re$  $(Im)$ denote the real (imaginary) part. In Fig.~\ref{FMean} we report the $\bar{F}_T$  for a chain with 
$N=22$ sites for different values of $g$ comparing it with the effective description 
$\bar{F}_{\text{eff}}$. It can be seen that already for values of $g=0.025 J$, the effective description is faithful and the average fidelity of teleportation is very close to unit at time $t^*$ obtained from the perturbative analysis (see inset of Fig.~\ref{FMean}).
Finally, let us mention that, decoupling the end spins from the chain, they are left in a state close to the maximally mixed state one. As $\comm{\hat{H}_{ij}}{ \hat{\rho}_{ij}}=0$, for $i,j=1,2$ and $N-1,N$, also 
the entanglement of teleportation stays almost constant, exhibiting oscillations less that $1\%$ of the values at $t=\frac{\pi}{2}$ in Fig.~\ref{EE.1}, as we have numerically verified. 

\section{Conclusions}\label{S.Concl}

We worked out a protocol for generating four-qubit generalized Bell states, to be used in 
quantum teleportation of an arbitrary two-qubit state, via the natural Hamiltonian dynamics of a XX spin-1/2 chain
with weakly coupled end blocks. We obtained analytically a reduced set of eigenstates that describes 
faithfully the full dynamics of the system in the two-excitation manifold 
up to second-order perturbation theory. We found that a simple initialisation of the sender and the receiver blocks, i.e., a two-spin flip on a overall fully polarized spin background of the quantum channel, results in the generation of
the appropriate resource (entanglement) upon which the teleportation protocol will rely on. 

Considering the need in several quantum information processing tasks to transfer, with a minimal protocol, an $n>1$ qubit state, we have set the first steps in this direction by implementing the case $n=2$ in a quantum channel which fulfils also the $n=1$ case. Remarkably, the time scale of sharing a tensor product of two Bell states is the same as that of sharing a single Bell state, hinting towards the possibility that the generation of a $n>2$ tensor product of Bell states is independent of the number of Bell pairs aimed to be generated. This seems to be a consequence of the non-interacting nature of the model, where the $n$-particle dynamics can be decomposed into one-particle ones and will be addressed in a further work.

Our work was inspired on the idea of using preengineered spin chains for transmitting (and generating) states from one point to another with minimal control, which may find applications in intermediate-scale quantum computations as well~\cite{Preskill2018quantumcomputingin}. Further extensions of this work should generalize
to protocol to cover the generation of resources for $n-$qubit teleportation
as well as investigate the effects
static disorder and other forms of noise, as well as other ways to perturbatively couple the sender and the receive blocks to the quantum channel ---\textit{e.g.}, strong local magnetic fields~\cite{qst1, qst2, qst4}.
Considering, finally, the high level of control achievable in cold atoms settings, we believe that our protocol is within experimental reach \cite{Volosniev2015,Khajetoorians1062,Banchi2017}.

\bibliographystyle{apsrev4-1}
\bibliography{biblio}

%\onecolumn\newpage
%\appendix

%\section{First section of the appendix}

\end{document}